\documentstyle[12pt]{article}

\begin{document}

\newcommand{\beq}{\begin{equation}}
\newcommand{\eeq}{\end{equation}}

\newcommand{\beqn}{\begin{eqnarray}}
\newcommand{\eeqn}{\end{eqnarray}}

\newcommand{\ra}{\rightarrow}

\newcommand{\su}{$ SU(2) \times U(1)\,$}

\newcommand{\nn}{\noindent}
\newcommand{\la}{\lambda}

\newcommand{\np}{Nucl.\,Phys.\,}
\newcommand{\pl}{Phys.\,Lett.\,}
\newcommand{\pr}{Phys.\,Rev.\,}
\newcommand{\prl}{Phys.\,Rev.\,Lett.\,}
\newcommand{\prep}{Phys.\,Rep.\,}
\newcommand{\zp}{Z.\,Phys.\,}

\newcommand{\eps}{\epsilon}
\newcommand{\mw}{M_{W}}
\newcommand{\mww}{M_{W}^{2}}
\newcommand{\mz}{M_{Z}}
\newcommand{\mzz}{M_{Z}^{2}}

\newcommand{\epm}{$e^{+} e^{-}\;$}
\newcommand{\gag}{$\gamma \gamma \;$}
\newcommand{\ggww}{$\gamma \gamma \ra W^+ W^-$}
\begin{titlepage}
\pagestyle{empty}
\rightline{July 1993}
\rightline{ENSLAPP-A-430/93}
\rightline{hep-ph/9308342}

\begin{center}

\vskip 1.cm

{\Large{\bf Higgs or Neutral Vector Boson Production with a $W$ Pair in
{\LARGE $\gamma \gamma$} Collisions}}

\vspace*{2cm}

{\large{\bf M. \ Baillargeon}}\footnote{On leave from
{\em Laboratoire de Physique Nucl\'eaire, Universit\'e de Montr\'eal,
C.P.  6128, Succ. A, Montr\'eal, Qu\'ebec, H3C 3J7, Canada.}}
and {\large{\bf F. \ Boudjema }}
          \vspace{0.5cm} \\
{\it Laboratoire de Physique Th\'eorique}
EN{\large S}{\Large L}{\large A}PP
\footnote{URA 14-36 du CNRS, associ\'e \`a l'E.N.S de Lyon, et au L.A.P.P.
d'Annecy-le-Vieux.}\\
{\it Chemin de Bellevue, B.P. 110, F-74941 Annecy-le-Vieux, Cedex, France.}

\end{center}

\vspace*{2cm}

\begin{abstract}

\nn Exploiting the fact that $W$ pair production in
high-energy $\gamma \gamma$ collisions is very large,
we use this process to trigger Higgs, $Z$ or photon radiation.
We find that there are sizeable rising cross-sections for triple bosons
production. At
energies about $1TeV$ the
new mechanism for Higgs
production becomes very competitive with the dominant
Higgs production
processes in $e^+e^-$ and $e \gamma$ reactions. The
effect of different polarized photon spectra obtained through
back-scattered laser light on the electron beam of a linear collider is
investigated .  We give
a special attention to the search of the intermediate mass Higgs in $WWH$
production and discuss how to effectively suppress the backgrounds.

\end{abstract}

\end{titlepage}

\newpage

\section{Introduction}
The ongoing intense activity in the physics potential offered by a
high-energy \epm linear collider has stimulated a growing interest in
the possibility of turning such a machine into a high-energy and
high-luminosity
$\gamma \gamma$ collider\cite{ggcol}. The large flux of very energetic photons
is obtained through Compton backscattering of laser light on the single-pass
electrons of the linear collider. Some of the principal attractions of running
in this mode rest on the fact that this is a `` democratic" means of
producing all
charged particles and that neutral scalar particles, notably the Higgs, can be
produced as a resonance in the $s$-channel. The search for the Higgs in this
mode has in fact gathered most attention. \\
\noindent For a TeV (or even a mid-TeV) \epm collider the $\gamma \gamma$ mode
is a more efficient way of producing $W$ pairs. Indeed, the cross section
for the process $\gamma \gamma \ra W^+ W^-$ is very large and
does not decrease at high-energy due the spin-1 t-channel $W$
exchange\cite{nousggvv}. Already at an effective
$\sqrt{s}_{\gamma \gamma} \sim 400GeV$ the reaction reaches its
plateau with a cross section of $\sim 80 pb$. The asymptotic {\em constant}
cross-section is
$\sigma_{\rm asymp.}(\gamma \gamma \ra W^+W^-) \sim 8\pi \alpha^2/M_W^2$.
This is more than an order of magnitude
larger than $W$ pair production in the usual $e^+ e^-$ mode at the same
centre-of-mass energy. The latter, as is known, decreases with energy,
$\sigma_{\rm asymp.}(e^+e^-\ra W^+W^-) \sim (\pi \alpha^2 s_W^{-4}/2 s)
\log(s/M_W^2) $.
This also means that the currently
discussed Next Linear $e^+ e^-$ Collider operating at $500GeV$ with a yearly
luminosity of $10-20fb^{-1}$ will produce about one million $W$-pairs in the
$\gamma \gamma$ mode. One may then even contemplate using this reaction as
a luminosity monitor. In this letter we exploit this process as a backbone
reaction to which we ``graft" one more additional boson. We will show that
this reaction triggers a sizeable Higgs cross section and that $WWZ$ and
$WW\gamma$ productions are even larger.
\section{The use of a non-linear gauge at tree-level}

The processes contributing to the triple boson production in \gag
are shown in Fig. 1. Because of the relatively large
number of diagrams an efficient way of calculating is almost mandatory.
Calculating in the usual unitary gauge is rather awkward because of the
cumbersome presence of the ``longitudinal" mode
(``$k_\mu k_\nu$" term) of the various W propagators.
A way out is to use a Feynman gauge. However, with the widespread choice
of a linear gauge fixing term,
this is done at the
expense of having to deal with even more diagrams containing the unphysical
Higgs scalars. An indisputable choice of gauge for photonic reactions or
for processes involving a mixture of W's and photons is to quantize with a
non-linear gauge fixing term \cite{nonlinear}
and work with a parameter corresponding
to the
't Hooft-Feynman gauge. For the processes at hand this means that one has
the same number of diagrams as in the usual unitarity gauge save for the fact
that we have no ``longitudinal" mode to worry about and that there are no
diagrams
with unphysical scalars since the virtue of this choice is that
the vertex with the photon, the $W$ and the unphysical Higgs field
does not exist.
Of course, one has to allow for small changes in the vertices which turn out
to have an even more compact form than in the usual gauges. For instance,
all diagrams where the quartic $WW\gamma \gamma$ vertex appears are identically
zero when the two incident photons have opposite helicities ($J_Z=\pm 2$).
We foresee this choice to stand out for applications
to W dynamics at a future $\gamma \gamma$ collider as it has proved to be
for one-loop weak bosons induced amplitudes for photonic processes
\cite{Z3g}.  \\

\nn With  $S^{\pm}$ being the unphysical Higgs bosons,
the $W^\pm$-part of the linear gauge fixing condition

\beq
{{\cal L}}^{Gauge-Fixing}_{linear}\;=\;-\xi^{-1}
|\partial_\mu W^{\mu +}\,+\,i\xi M_W S^{+}|^{2}
\eeq

%$A^{\mu} W^{\pm}_{\mu} S^{\pm}$ vertex.\\

\nn is replaced by the ``constraint"

\beq
{{\cal L}}^{Gauge-Fixing}_{non-linear}\;=\;-\xi^{-1}
|(\partial_\mu\;+\;i e A_\mu\;+\;ig \cos \theta_W Z_\mu) W^{\mu +}\,+\,
i\xi M_W S^{+}|^{2}
\eeq

\nn where $\theta_W$ is the usual weak mixing angle. We have taken $\xi=1$.

\nn Having reduced the number of diagrams by the gauge-fixing choice we have
eased our computational task by calculating the full helicity amplitudes.
These were, in turn, fed into a Monte-Carlo event generator which preformed
the phase space integrals.
We have checked that our amplitudes were gauge invariant both analytically
and numerically. We will take $M_W=80.1$GeV, $M_Z=91.18$GeV and
$\sin^2 \theta_W=0.232$.

\section{Behaviour of the cross sections}
We first present our results for an ideal $\gamma \gamma$
collider to explicitly exhibit the interesting behaviour of the various
cross sections with \gag centre-of-mass energy. We will then include the
effect of more realistic luminosity spectra.

\subsection{$\gamma \gamma \rightarrow W^+ W^- \gamma$}
For the $WW\gamma$ final state, a cut on the ( final) photon energy is
required. One may also prefer to take a cut on the transverse momentum of
the photon. With a fixed cut $p_T^\gamma > 20$GeV for all centre-of-mass
energies, the cross section increases with energy.
At $500$GeV one reaches a
cross section of about $1.3$pb. This is about $1.6\%$ of the
$WW$ cross section at the same energy.
The $J_Z=0$ obtained when both photons have the
same helicity slightly dominates over the $J_Z=2$ ($1.5$pb versus $1.1$pb).
At $\sqrt{s}_{\gamma \gamma}=2$TeV the cross section with the same
$p_T^\gamma >20GeV$ cut reaches $3.7$pb. The logarithmic ($\log^2 s$) growth
can be understood
on the basis that this cross section can be factorised in terms
of $\gamma \gamma \ra WW$, which is constant at asymptotic $M_{WW}$
invariant masses, times the final state photon radiator which contains
the logarithmic $s$ dependence. We note that this logarithmic
increase only concerns the production of transverse $W$. When both
$W$ are longitudinal ($W_L W_L$) the cross section decreases. This can also
be traced back to the fact that $\gamma \gamma \ra W^+_L W^-_L$ decreases
with energy. We find that the $W_L W_L$ fraction of all $W$'s is about
only a $1\%$ at $500$GeV, $0.3\%$ at $1$TeV and a mere $0.07\%$ at $2$TeV.
It must be noted that the bulk of the cross section occurs when all
final particles are produced at very small angles: this is a typical example
of multiparticle production in the very forward region. For instance increasing
the $p_T^\gamma$ cut and at the same time imposing a pseudorapidity cut
on the photon, the $WW\gamma$ yield, as shown in Table 1., drops
considerably, especially at higher energies. The reduction is even more
dramatic when
we put an isolation cut between all the particles and forcing them away from
the beam. With these strictures the cross section decreases with energy
(See Table 1.)

\begin{table}[here]
\begin{center}
\begin{tabular}{|l|c|c|c|c|}
\hline
$\;\;\;\;\;\;\;\;\;\;\;\;\;\;\;\sqrt{s}_{\gamma \gamma}$(TeV) &0.5&1&1.5&2 \\
%\cline{2-5}
&&&&\\
type of cut&&&& \\
\hline
1. &1254&2469&3195&3678\\
2. &1254&1434&1258&1050\\
2. and 3.&1235&1373&1159&930 \\
2. and 4.&201&86&47&32 \\
\hline
\end{tabular}\\
\vspace{0.4cm}
\parbox{10.5cm}{{\small{1. $p_T^\gamma >$ 20 GeV  \hspace*{1.12cm}
2. $p_T^\gamma >40$GeV$\times\sqrt{s}_{\gamma \gamma}$(in TeV) \\
3. $|y^\gamma|<2$ \hspace*{1.99cm}
4. $\cos$(between any two particles)$<0.8$}}}
\end{center}
{\bf Table 1:} {\footnotesize{\em Cross section for
$\gamma \gamma \rightarrow W^+ W^- \gamma$ (in fb) at different
$\sqrt{s}_{\gamma \gamma}$ including various cuts.}}
\end{table}

\noindent While the $W_L W_L$ production is very much favoured
in the $J_Z=2$ mode, $W_TW_T$ and $W_T W_L$
productions (which are by far the largest
contributions) are slightly more favoured in the $J_Z=0$ channel (see Fig.~2).
This is the same behaviour as in the two body process \ggww.

%WWZ  WWZ WWZ  WWZ

\subsection{$\gamma \gamma \rightarrow W^+ W^- Z$}
\nn Contrary to the previous reaction one can calculate the
total cross section. It exhibits an interesting behaviour
at TeV energies.
One notes that already
at $1$TeV the triple vector boson production is larger that top pair
($m_t \geq 130$GeV)
and charged heavy scalars production (with $m_{H^{\pm}}\simeq 150$GeV) as
shown in Fig. 3 which compares various process in $\gamma \gamma$ and
$e \gamma$ collisions.
At $2$TeV the total $WWZ$ cross section is about $2.8$pb
and exceeds the
{\em total} electron-positron pair production. This is a typical example
of the increasing importance of multiparticle production in weak interactions
at higher energies, purely within perturbation theory
\footnote{As opposed to the hypothetical surmise of large $W$ multiplicities
due
to topological effects at extremely high-energies.
For a recent review see \cite{Ring}.}. The rising of the cross-section
with the centre-of-mass energy is essentially from the very forward region
due to the presence of the ``non-annihilation" diagrams with the (spin-1)
$W$ exchanges. A similar behaviour in \epm reactions is single vector
boson production. What is certainly more interesting in
$\gamma \gamma \ra W^+W^-Z$ is the fact that it has a purely
non-abelian origin.
It may be likened to $gg \ra ggg$ in QCD except that we do not
need any infrared cut-off (the $W$ and $Z$ mass provide a natural cut-off).

\noindent
The bulk of the cross section consists of both $W$ being
transverse as is the case with the ``parent" process
$\gamma \gamma \ra W^+W^-$.
While the total cross-section is larger in the $J_Z=0$ than in the $J_Z=2$,
the production of all three vector bosons being longitudinal occurs mainly
in the $J_Z=2$ channel and accounts for a dismal contribution. For instance,
the ratio of $LLL/TTT$ (three longitudinal over three transverse) in the case
of unpolarized beams amount to a mere $2$per-mil at $500$GeV and drops to
$0.1$per-mil at $2$TeV.
Nonetheless, the total \footnote{i.e., taking into account all polarization
states of the $W$.}  production of longitudinal $Z$'s
as compared to that of transverse $Z$
is not at all negligible. In fact, between $500$GeV and $2$TeV this ratio
increases from about $23\%$ to $32\%$ (see Fig. 4).
This is somehow counterintuitive
as one expects the longitudinal states to decouple at high energies.
The importance of $Z_L$ production (in association with $W^+_T W^-_T$) is,
however, an ``infrared" rather than an ``ultraviolet" phenomenon in this
reaction: the $Z$ is not energetic.
First, one has to realize that the
$\gamma \gamma \ra  W^+W^-Z$ amplitude
is transverse in the momentum of the $Z$, $q$, as is the case with the photons
in $\gamma \gamma \ra W^+W^- \gamma $.
With $k_1$ and $k_2$ being the momenta of the photons, the longitudinal
polarization vector of the $Z$, with energy $E_Z$, writes

\beq
\epsilon_{\mu}^L =\frac{1}{\sqrt{E_Z^2-M_Z^2}} \left(\frac{E_Z}{M_Z}q_\mu
\;-\; \frac{M_Z}{\sqrt{s}} (k_1+k_2)_\mu \right) \; ; \;\
\epsilon^L.\epsilon^L =-1
\eeq

\noindent
The transversality of the amplitude means
that the leading (``ultraviolet" $\propto E_Z$) part does not contribute.
Only the
``infrared" part $\propto M_Z$ does.
This contribution should vanish in
the limit of vanishing $M_Z$.
However, the amplitude, in analogy with what happens in $WW\gamma$, has the
infrared factor
$1/E_Z$ and the ``soft" term in equation (3) contributes.
Furthermore, more importantly, the bulk of
the cross section is from configurations where both $W$ are transverse
(see Fig.~4) and
all three particles go down the beam. In the limit of
vanishing masses this topology leads to collinear divergences
\footnote{In the limit $M_V \ra 0$,
added to the divergence in $\gamma \gamma \ra W^+W^-$, there is
the collinear divergence when the $Z$ and a $W$ are collinear.}.
In this dominating configurations, in the exact forward direction,
the longitudinal $Z$
contributes maximally. At the same time, angular momentum
conservation does not allow the $Z$ to be transverse when all final
particles are down the beam (with $p_T=0$) and both $W$ are transverse.
So the ``maximal collinear enhancement" is not as operative for the
transverse $Z$
as it is for longitudinal $Z$ when both $W_T$ are at zero $p_T$.
However, as soon as one moves away from these singular configurations, the
longitudinal $Z$ does decouple and the ``smooth" mass limit may be taken.
This is well rendered in Table.2
which displays the ratio of $Z_L/Z_T$ without any cut and with the inclusion
of cuts. The most drastic of these cuts is when we impose angular separation
cuts between the final particles and
forcing them to be away from the beam, with the effect that
the $Z_L/Z_T$ decreases with energy and gets dramatically smaller.

\begin{table}[here]
%\begin{center}
\begin{tabular}{|l||c|c||c|c||c|c||c|c||}
\cline{2-9}
\multicolumn{1}{c||}{} &\multicolumn{2}{c||}{$\sqrt{s}_{\gamma \gamma}=500$GeV}
&\multicolumn{2}{c||}{$\sqrt{s}_{\gamma \gamma}=1$TeV}
&\multicolumn{2}{c||}{$\sqrt{s}_{\gamma \gamma}=1.5$}
&\multicolumn{2}{c||}{$\sqrt{s}_{\gamma \gamma}=2$TeV} \\
\multicolumn{1}{c||}{}
&$\sigma$(fb)&L/T &$\sigma$(fb)&L/T &$\sigma$(fb)&L/T &
$\sigma$(fb)&L/T \\
\hline
no cut&428&$24\%$&1443&$27\%$&2195&$30\%$&2734&$32\%$ \\ \hline
$\cos(WZ)<0.8$&368&$19\%$&1025&$18\%$&1321&$19\%$&1465&$19\%$ \\ \hline
$\cos(W\;{\rm beam})<0.8$&164&$18\%$&232&$17\%$&186&$17\%$&145&$16\%$ \\ \hline
$\cos(\;{\rm ``all"})<0.8$&115&$11\%$&140&$8\%$&105&$6\%$&77&$5\%$ \\ \hline
$E_Z>150$GeV&184&$11\%$&1032&$21\%$&1700&$25\%$&2220&$28\%$ \\ \hline
\end{tabular}
\vspace*{0.3cm}
{\bf Table 2:} {\footnotesize{\em Cross section for
$\gamma \gamma \rightarrow W^+ W^- Z$ and ratio of longitudinal over
transverse $Z$ (L/T) including various cuts. ``all" means that we require
the final particles to be separated \underline{and} to be away from the beam
by an angle $\theta$ corresponding to $\cos \theta <0.8$.}}
\end{table}

\noindent A more detailed study with exact analytical expressions for the
helicity amplitudes exhibiting the above behaviour is left for a longer
publication \cite{nousgg3}.

\noindent
On the phenomenological side, the study of this reaction is important
as, especially for $M_H \sim M_Z$, it is a background to Higgs detection
through $WWH$ production  to which we now turn.

\subsection{$\gamma \gamma \rightarrow W^+ W^- H$}

\nn While it is almost certain that the LHC/SSC will discover a Higgs if
its mass is above $2M_Z$, the intermediate mass Higgs, {\em IMH},
will be extremely
difficult to track at these machines. This ``mass gap"
will
be efficiently covered by the next $e^+e^-$ linear collider where two
complementary reactions are at work: the so-called Bjorken process
$e^+e^- \rightarrow ZH$ dominating at moderately low energies and
the $WW$ fusion process which starts to dominate above $\sim 500GeV$
\footnote{For a recent comparison between the different modes of Higgs
production in \epm see, \cite{eevvh}.}.
One of the original motivations for a $\gamma \gamma$ collider
is to produce the scalar Higgs as a resonance. To set the stage,
let us
recall that, in the range: $90 <M_H<140 GeV$,
and considering
the dominant decay of the Higgs into b-quarks, the cross section
$\sigma({\gamma \gamma \stackrel{H}{\rightarrow} b \bar{b} } )$ is about
$\sim 50-60fb$, for an
optimal set of cuts and parameters of the $\gamma \gamma$ collider
\cite{Borden} . \\
%\begin{figure}[here]
%\vspace{9.5cm}
%$\caption{Comparison of 4 standard Higgs production:
% $e^+ e^-\rightarrow \bar{\nu}\nu H$,
%$e^+ e^-\rightarrow ZH$ and $e \nu \rightarrow W H \nu$ at 500 GeV and 1 TeV.}
%\label{higgs}
%\end{figure}

\noindent
Another efficient mechanism for Higgs production in an $e \gamma$
environment is through $e \gamma \ra \nu W H$ \cite{egnuwh}. Still, in
the context of $\gamma \gamma$ collisions, it has recently been suggested
\cite{ggtth}
to look at the production of Higgs in association with a top pair in analogy
to $t \bar t H$ production in hadron machines. Unfortunately, the {\em IMH}
yield
does not exceeds $1-3fb$ (for $m_t\leq 150$GeV). The $t\bar t H$ cross section
decreases very slowly with $\sqrt{s}_{\gamma \gamma}$. \\
%crossand the cross section decreases with energy as
%does the ``sub-process" $\gamma \gamma \ra t \bar t$.\\
\noindent
Taking, once again, advantage of the large $WW$ cross section, we propose to
search for Higgs in association with a $W$ pair.
We find that for a Higgs mass
of $100$GeV we obtain a cross section of about $20$fb at
$\sqrt{s}_{\gamma \gamma}=500GeV$.
%Although at this energy one is penalized by
%a rather narrow phase space, the associated $W$ production is by an order of
%magnitude larger than the corresponding associated top production.
The $WWH$ cross section quickly rises to yield $\simeq 400fb$ at
$2$TeV (for $m_H=100GeV$). The importance  of this mechanism
at TeV energies is best illustrated, by contrasting it with top pair
production (see Fig. 3).
For $m_H=100$GeV and $m_t=130$GeV (consistent with present
{\em indirect} LEP limits), the two process have the same threshold energy
and lead to the same final state ({\em IMH}
decays predominantly into $b \bar b$).
While at $\sqrt{s}_{\gamma \gamma} \simeq 500$GeV top pair production is
almost two-orders of magnitude larger than $WWH$, the latter which is a third
order process is twice as large at $2$TeV. Nonetheless, the $WWZ$
cross section is about an order-of-magnitude larger than the ``{\em IMH}-$WWH$"
for all centre-of-mass energies.
%Phenomenologically, one should pay a
%special attention to $WWZ$ when looking for Higgs in $WWH$.
%But, if $b$-tagging
%is available one should be able to bring down the $WWZ$ background to
%the $WWH$ signal since the $B(Z \ra b \bar b) \simeq 15\%$.

\noindent Comparing at the \underline{same}
$\sqrt{s}_{\gamma \gamma}$ and $\sqrt{s}_{e \gamma}$ centre-of-mass,
in the {\em IMH} case,
the cross sections for $WWH$ start becomming larger than those
of $e \gamma \ra \nu W H$ for energies around $700$GeV. At lower energies
the $e\gamma$ mode benefits from a larger phase space (see Fig. 3). \\
\noindent
In Fig. 5 we contrast the various mechanism of Higgs production in an $e^+e^-$
environment in the \epm, \gag, and $e \gamma$ modes before folding with the
luminosity spectra. In the {\em IMH} case, taking for illustration $M_H=80$GeV,
at $500$GeV, $\sigma(\gamma \gamma \ra WWH)\simeq 30$fb which is
by only factor $2$ smaller than $\sigma(e \gamma \ra \nu WH)$ and a factor
$3.3$ compared to the dominant $WW$ fusion process in $e^+e^-$.
On the other hand, $\sigma(\gamma \gamma \ra WWH)$
is larger than all the $VVH$ ($WWH, ZZH, ZH\gamma)$ processes
in \epm
by at least a factor $3$. Higgs production from top bremstrahlung ($t\bar{t} H$
final state),
either in \epm \cite{eetth} or \gag \cite{ggtth}
is abysmally small. At $1TeV$
our process becomes very comparable to $e\gamma \ra \nu WH$ and is only
about a factor 2 smaller than the dominant
$WW$ fusion process in \epm.
Nonetheless, the fact that in $\sigma(\gamma \gamma \ra WWH)$,
unlike the $WW$ fusion in $e^+ e^-$ or the corresponding one at
$e \gamma$, all final particles
can be observed or reconstructed
(hence alleviating the lack in
energy constraints) makes this reaction worth considering especially at a
TeV $\gamma \gamma$ collider. But of course, this statement tacitly assumes
an ideal monochramatic \gag collider. We will now turn
to more realistic photon luminosity spectra. Before so doing , it is
worth pointing out that an almost equal number of $H$ is produced in the
$J_Z=0$ or the $J_Z=2$ with both $W$ being essentially transverse.
%$at 500~GeV, $\gamma \gamma \rightarrow W^+ W^-H$ is rather small
%due mainly to a lack of
%phase space for the 3 massive final particules even if for
%$M_H=90GeV$ one has $10fb$ which is a third of the cross section
%for $e \gamma \rightarrow \nu W H$ (see Fig.2) .
%However, at higher energies this mechanism becomes important.
%For instance at 1TeV and for Higgs masses all the way up to $500GeV$ the
%cross section is larger than that from the Bjorken process (see Fig. 2)
%and almost equal to that in the $e \gamma$ mode, although somehow
%still lower than with the fusion mechanism. For instance, the cross section
%for a $150GeV$ Higgs at 1TeV (800 GeV for $\gamma \gamma$) is about 90fb.
%Considering a typical yearly luminosity of $60 pb^{-1}$ at these energies
%this mechanism will produce about 5400 Higgs. Of course, one expects some
%reduction in this number when appropriate cuts and branching ratios are taken
%into account.
\section{Inclusion of the photon luminosity spectrum}
Thus far, two-photon processes at $e^+e^-$ have exploited the
``Weisz\"acker-William"
spectrum, which is essentially a ``soft-photons" spectrum. The $\gamma \gamma$
luminosity peaks for very small fraction of the invariant $\gamma \gamma$ mass
$\sqrt{s}_{\gamma \gamma}$, i.e., for
$\tau=s_{\gamma \gamma}/s_{e^+e^-} \ll 1$. The laser scheme on
the other hand permits to transmit a very large proportion of
the energy of the
electron ($E_e$)
to the ``collider" photon by shining a low-frequency ($\omega_0$) laser beam
at a glancing angle. With $x$ being the reduced invariant mass
of the original $e\gamma_{{\rm laser}}$ system:
$x\simeq 4E_e \omega_0/m_e^2$, the maximum
energy fraction, $y_{max}=\omega_{max}/E_e$, that the colliding photon can take
is $y_{max}=x/(x+1)$. This occurs for photons produced in
the exactly forward ($e^-$) direction.
One then has to tune the energy of the laser so that one gets the highest $x$.
However, this $x$ can not be arbitrarily large, otherwise one reaches the
threshold for $e^+e^-$ creation by the interaction of the laser beam and the
converted photon. This occurs for $x=x_0\simeq 4.8$ \cite{ggcol}
and translates into
$\sqrt{\tau}_{|max} \approx 0.83$. Our analysis is based on
taking this value of $x_0$ for all $e^+e^-$ energies,
which means that one has to take
different laser frequencies for different $\sqrt{s}$ colliders. \\
\noindent
To achieve a higher degree of monochromaticity of the spectrum, polarization
is
essential. Instead of writing uninspiring lenghty formulae for the polarized
luminosity spectra, we prefer to refer to Fig.~6 (see also \cite{ggcol}).
It shows that the hardest
spectrum is arrived at
by choosing the circular polarization
of the laser ($P_c$) and the mean helicity of the electron ($\lambda$)
to be opposite. For the photon mode of the collider this means
$2 \lambda P_c=2 \lambda' P'_c=-1$
($'$ are for the opposite arm of the photon collider).
Fig.~6 shows the case where both lasers
are tuned to have a right-handed circular polarization ($P_c=P_c'=+1$).
This has the added advantage that the high-energy photons are produced
mostly with the same helicity therefore giving a $J_Z=0$ dominated
environment, for short we will refer to this as the ``0-dom." case.
The $J_Z=2$ tail almost disappears for $\sqrt{\tau}>0.7$.
For some processes where the $J_Z=2$ is dominant,
or if one wants to compare the $J_Z=2$ and the $J_Z=0$ on an
``equal basis", one
would also like to isolate the $J_Z=2$ at the expense of the $J_Z=0$ spectrum.
We point out that this could be easily achieved by flipping {\em both} the
electron and laser polarizations of {\em one} of the arms {\em only} while
maintaining $2 \lambda P_c=-1$ (for a maximum of monochromaticity). In this
case,
 the  $J_Z=0$ and  $J_Z=2$ spectra in Fig.~6 have to be interchanged, for
short ``2-dom.". The
spectrum one expects in case of no polarization is rather flat, with a slight
hump in the ``mid-range" $\sqrt{\tau}\simeq 0.2-.5$. For processes
which increase with energy, as with the three processes we have studied,
it is best to choose the hardest spectrum arrived at through oppositely-handed
$e, \gamma_{{\rm laser}}$. This also helps in sensibly reducing standard
processes which at the ``partonic" ($\gamma \gamma$) level drop as $1/s$.
We will discuss the effect of the luminosity spectrum in the reactions we have
studied for the case of polarized beams and in the case of no polarization.
%In the former, since we have seen that there is not,especially at
%high energies, a clear-cut preference between the $J_Z=0$ and the $J_Z=2$
%we expect to obtain similar results as long as
%$\lambda P_c=\lambda' P'_c=-1$.

\section{Folding with the luminosity spectra}
We illustrate the effect of different luminosity spectra by
concentrating on the {\em IMH}
search through $WWH$. This will lead us to consider
$WWZ$ production which is the most obvious background for $M_H \sim M_Z$.
With $\sqrt{s}_{ee}=500$GeV, the inclusion of the spectra changes
the $WWH$ yield
significantly due to the fact that the mamixum
$\sqrt{s}_{\gamma \gamma} \simeq 400$GeV  leaves a small phase space for the
{\em IMH}.
Even when we choose the polarization of the primary beams to give the
peaked $J_Z=2$-dominating spectrum (``2-dom".),
the cross section does not exceed $4$fb and
is therefore almost two orders of magnitude below the $WW$ fusion process in
the \epm mode and an order of magnitude smaller than $\nu W H$ production in
the
$e \gamma$ mode (See Fig.~7a).
The situation is much more favourable at $1$TeV. Up to
$M_H \simeq 300$GeV this mode produces almost twice as many Higgses as the
conventional Bjorken process. For $M_H=100$GeV and choosing a setting
which gives a ``0-dom", we obtain $\sim 37.5$fb (compared to $37.2$ in
the ``2-dom") and
$26.3$fb with no polarization for the primary beams.
The advantage of a polarized spectrum is undeniable. $\nu W H$ (in $e \gamma$)
and $H \nu \nu$ (in \epm) are respectively about 2 and 5 times larger in the
{\em IMH}
case. A comparison between the variety of Higgs production modes in the
NLC(1TeV) environment is shown in Fig.~7b which clearly brings out the
importance of $WWH$.

\vspace{0.3cm}
\begin{table}
\begin{tabular}[tbh]{|l||c|c|c||c|c|c||}
\cline{2-7}
\multicolumn{1}{l||}{}&\multicolumn{3}{c||}{500GeV}&\multicolumn{3}{c||}{1TeV}
\\ \cline{2-7}
\multicolumn{1}{l||}{}&non pol.&0-dom.&2-dom&non pol.
&0-dom.&2-dom\\ \hline
$WWH$&1.0&1.7&2.24&26.3&37.5&37.2 \\
$WWH_{\hookrightarrow b \bar b}$&0.8&1.4&1.8&21&30&29.8\\
$WWH_{\hookrightarrow b \bar b}$ ``non-top"&0.7&1.1&1.5&20.2&28.9&28.7 \\
$WWZ$&24.2&55.3&36.9&342&473&408 \\
$WWZ_{\hookrightarrow b \bar b}$&3.6&8.3&5.53&51.3&70.9&61.2 \\
$WW\gamma$&205&321&272&483&592&560\\
$t \bar t$ (``direct")&207&458&250&620&525&687 \\
\hline
\end{tabular}
\vskip .3cm
{\bf Table 3}:
{\footnotesize {\em Cross section in fb for three-boson (and $t \bar t$)
productions with $M_H=100$GeV and $m_t=150$GeV. The $WW\gamma$ includes
a $p_T^\gamma$ cut of $20$GeV at
$500$GeV and $40$GeV at $1$TeV. ``non-top" means all Higgs events with Higgs
decaying into $b \bar b$ and where the simulataneous $Wb$ invariant mass has
been applied as explained in the text. ``direct" means that we have not
taken into account top pairs produced through the gluons inside the photon,
i.e., the ``resolved" photons contribution has not been considered.}}
\end{table}

\noindent
The effect of switching between different polarization settings is even
more drastic in the case of $WWZ$. The largest cross sections are in
the ``0-dom." case. At $\sqrt{s}_{ee}=500GeV$ we find
$\int \sigma(WWZ) \sim 55fb$ which is twice as large as the non-polarized
case (24.2fb). Note that, when choosing the ``0-dom" the $WWZ$ yield is larger
than in the conventional \epm
mode ($\sigma(e^+e^- \ra W^+W^-Z)_{|\sqrt{s}_{ee}=500{\rm GeV}}$ $\sim 40fb$)
\cite{quarticplb}. At $1$TeV the $WWZ$ reaches $\sim 470$fb in the ``0-dom"
and is slightly smaller ($410$) in the ``2-dom". \\
\noindent
Considering the large $WWZ$ yield, $b$ tagging is almost necessary for the
{\em IMH}
search. Another dangerous background, even for the case of $b$-tagging is
due to top pair production: $\gamma \gamma \ra t \bar{t} \ra W^+W^- b \bar{b}$.
For instance, at $\sqrt{s}_{ee}=500GeV$ this is about two-orders of magnitude
larger
than $WWH_{\hookrightarrow b \bar b}$. Fortunately, one can eliminate this huge
contamination by rejecting all those $WWH$ events where
the \underline{simultaneous} cuts on the invariant mass of the two $Wb$ is
such that the $Wb$ does not
reconstruct the top mass (within $15GeV$)

\beqn
m_t - 15GeV < M_{W^+b} < m_t + 15GeV \;&{\rm and}&\;
m_t - 15GeV < M_{W^{-}b'} < m_t + 15GeV \nonumber \\
&{\rm or}& \nonumber \\
m_t - 15GeV < M_{W^+b'} < m_t + 15GeV \;&{\rm and}&\;
m_t - 15GeV < M_{W^-b} < m_t + 15GeV \nonumber \\
\hspace*{2cm}
\eeqn

\noindent
The reason we try both combinations $W^+b$ or $W^+b'$ is that we do not want
to rely on charge identification, for the $b$ especially, which necessarily
entails a reduction in the $b$ sample (and hence our signal). A good vertex
detector should be sufficient
\footnote{We have not tried to cut the $t \bar t$ by demanding that
$m_{b\bar b}=M_H\pm10GeV$, as the cut above is very efficient. Moreover,
based on our previous analysis of $WWH$ in \epm \cite{eevvh}, the $Wb$ cut
was by far more efficacious.}. In carrying the vetoing in our Monte-Carlo
sample we made the Higgs decay isotropically in its rest frame. The
effective loss at $500$GeV is about a mere $0.3$fb while at $1$TeV, where
we have a ``healthy" cross section, the percentage loss is only about $4\%$
for all choices of the polarization. Table 3. shows the cross sections taking
a Higgs mass of $M_H=100$GeV with $Br(H \ra b \bar b) \sim 80\%$ and the cut
of equation 4,
assuming $m_t=150$GeV. Once the ``faked" top events have been dealt with,
the $WWZ_{\hookrightarrow b \bar b}$ do not bury the signal
(for $M_H \sim M_Z \pm 10$GeV) as
Table 3 shows. These $WWZ$ can be further reduced by judiciously switching the
``2-dom." setting, both at $500$GeV and at $1$TeV. Although at the former
energy
the event rate is probably too small to be useful, at $1$TeV, in the ``2-dom.",
we have, after including the cuts and the branching fractions into $b$, $30$fb
of signal compared to $60$fb from $WWZ$. With one $W$ at least, decaying into
jets and not taking into account decays into $\tau$'s, the number of $WWH$
with the contemplated integrated luminosity of ${\cal L}=60fb^{-1}$ will be
about 1400 events. Even if one allows for an overall efficiency of
$50\%$ this is a very important channel to look for the Higgs. There is one
background which we have not considered. It concerns the
$W^+W^- b \bar b$ final state with  $b \bar{b} \ra W^+ W^-$ as a
sub-process. We expect this to be very
negligible once one puts a high $p_T$ cut on both $b$'s and require
$m_{bb} \sim M_H$. We will give a more
detailed analysis of all these processes and a more thorough discussion
on background elimination in a longer forthcoming paper. \\

To conclude, we have shown that this new mechanism of Higgs production
in a $\gamma \gamma$ mode of $\sim 1TeV$ \epm collider
is a very promising prospect.
The oft discussed intermediate mass Higgs production, as a narrow
resonance in \gag collisions, relies on a spectrum which is peaked
around the Higgs mass in a $J_Z=0$ dominated setting. The extensive study
in \cite{Borden} finds that with $\int {\cal L}_{ee}=10fb^{-1}$, one expects
between about $500$ Higgs events for $M_H \sim M_Z$ to about $600$ events for
$M_H \sim 140$GeV. For the same
$\int {\cal L}_{ee}=10fb^{-1}$ this is about $2-3$ times more than what we get
with $WWH$ at $\sqrt{s}_{ee}=1$TeV. However, the resonance scheme means that
the available $\gamma \gamma$ invariant mass covers a very narrow, and
in the case of the {\em IMH}, low range of energies. Hence while allowing
a precise study of the $H\gamma \gamma$ coupling it forbids the
study of a wealth of interesting processes in the $\gamma \gamma$ mode of the
NLC. Higgs detection through $WWH$ at $1$TeV will be one aspect among a
variety of studies of weak processes ($WW, ZZ, WWZ$,...{\it etc}) in a
\gag environment.

\noindent
\vspace*{1cm}
\noindent {\bf \Large Acknowledgements}\\
\noindent We would like to thank Edward Boos, Misha Dubinin and Ilya Ginzburg
for fruitful discussions as well as confirming our result for $WWZ$ by
comparing it to the output of {\em CompHep}
in the case of unpolarized beams.

\newpage
\noindent
{\Large \bf Figure Captions} \\

\noindent
\underline{Fig. 1} Feynman graphs contributing to $\gamma \gamma \ra WWH, WWZ$
or $WW\gamma$. These have to be properly symmetrised. One then counts
4 diagrams of type (a) and (d), and 2 of types (b) and (c). \\
\noindent
\underline{Fig. 2} $\gamma \gamma \ra WW\gamma$ cross section as a function
of $\sqrt{s}_{\gamma \gamma}$ with $p_T^\gamma > 20$GeV, for different initial
and final polarization states. The subscript $_0$ and $_2$ refer respectively
to an initial state with $J_Z=0$ and $J_Z=\pm2$. TT is for both W being
transverse, TL when only one is transverse and LL when both are longitudinal.\\
\noindent
\underline{Fig. 3} Typical processes in $\gamma \gamma$ and $e \gamma$
reactions, with $m_t=130$GeV, $M_{H^{+}}=150GeV$. In $\gamma \gamma \ra
W^+W^-\gamma$ the cut is $p_T^\gamma >20$GeV. $e^+e^{-}_{\rm cut}$
represents
$\gamma \gamma \ra e^+ e^-$ with $|\cos (\gamma e)|< 0.8$.
The Higgs masses ($100$GeV and $200$GeV) are indicated by the subscripts
in $WWH$ and $WH\nu$. \\
\noindent
\underline{Fig. 4}  $WWZ$ cross section for different combinations of
final polarizations as a function of
$\sqrt{s}_{\gamma \gamma}$, with unpolarized photons. $Z_T$ ($Z_L$) is the
transverse (longitudinal) Z yield for any $W$ helicity state.
Also shown is the ratio
of longitudinal over transverse $Z$ ($Z_L/Z_T$) summed over all polarization
states of the $W$'s. \\
\noindent
\underline{Fig. 5} A comparison of Higgs production cross-sections in \epm
\gag and $e \gamma$ reactions at $500$GeV (5.a) and $1$TeV (5.b) before
folding with any photon luminosity spectrum. When the initial state is not
specified, it should be understood as an \epm process. \\
\noindent
\underline{Fig. 6} The $\gamma \gamma$ spectrum for different choices of the
primary electron longitudinal ($\lambda$) and photon circular
($P_c$) polarization with $x_0=4.82$. $'$ are for the opposite arm of the
collider.
The figure also shows the $J_Z=0$ and $J_Z=2$ part in the case of
$P_c=P'_c=+1$.\\
\noindent
\underline{Fig. 7} Comparison between different Higgs production mechanisms at
a $500GeV$ (a) and a future $1$TeV (b) \epm machine in the three modes of the
collider. No beam
polarization effects are included apart from $\gamma \gamma \ra WWH$
($WWH_{\rm pol}$) at $1$TeV
where we also  show the effect of a ``$J_Z=0$-dominated" setting (see text).


\begin{thebibliography}{99}
\bibitem{ggcol} I.F. Ginzburg, G.L. Kotkin, V.G. Serbo and V.I. Telnov,
{\em Nucl. Instrum. Methods} {\bf 205} (1983) 47;
I.F. Ginzburg, G.L. Kotkin, S.L. Panfil, V.G. Serbo and V.I. Telnov,
{\em ibid} {\bf 219} (1984) 5; V.I. Telnov, {\em ibid} {\bf A294} (1990) 72.
\bibitem{nousggvv}For a list of references on this process, see G. B\'elanger
and F. Boudjema, {\em Phys. Lett.} {\bf B288} (1992) 210.
\bibitem{nonlinear} M. Gavela, G. Girardi, C. Malleville and P. Sorba,
{\em Nucl. Phys.} {\bf B193} (1981) 257.
%{\em Phys. Lett.} {\bf B117} (1982) 64
\bibitem{Z3g}M. Baillargeon and F. Boudjema,
{\em Phys. Lett.} {\bf B272} (1991) 158 and references therein.\\
\hspace*{.5cm}A variant of the non-linear gauge fixing presented here
has been, very recently, been employed by G. Jikya for the calculation of
$\gamma \gamma \ra ZZ$, IHEP-preprint, IHEP-97/93.
\bibitem{Ring}A. Ringwald, CERN-preprint, CERN-TH-6862-93, April 1993.
%\bibitem{Morris}D.A. Morris and R. Rosenfeld, \np {\bf B391} (1993) 531.
\bibitem{nousgg3}M. Baillargeon and F. Boudjema, in preparation.
\bibitem{eevvh}M. Baillargeon et al., CERN-preprint, CERN-TH-6932/93,
  June 1993.
\bibitem{Borden}D.L. Borden, D.A. Bauer, D.O. Caldwell, SLAC-Preprint,
SLAC-PUB-5715/UCSB-HEP-92-01, January 1992. See also F. Richard,
Orsay-Preprint, LAL-91-62, Nov. 1991.
\bibitem{egnuwh} K. Hagiwara, I. Watanabe and P.M. Zerwas,
{\em Phys. Lett.} {\bf B278} (1992) 187. See also E. Boos et al., DESY-Preprint
91-114, Oct. 1991.
\bibitem{ggtth}E. Boos {\em et al.,} \zp {\bf C56} (1992) 487. \\
K. Cheung, \pr {\bf D47} (1993) 3750.
\bibitem{eetth}A. Djouadi, J. Kalinowski and P. Zerwas, \zp {\bf C54}, 255
(1992).
\bibitem{quarticplb}G. B\'elanger and F. Boudjema, \pl {\bf B288} (1992) 201.
\end{thebibliography}
\end{document}